# General relativity and the U(1) gauge group

Jean Paul Mbelek[(1)]

[(1)]Service d'Astrophysique, CEA-Saclay, Orme des Merisiers, 91191 Gif-sur-Yvette, France

October 03, 2009

**Abstract**

We show that gravity together with curved spacetime can emerge, at the microscopic scale, from a U(1) gauge field. The gauge boson that carries gravity, of elementary particles, is proved to be a spin one massless and electrically neutral vector particle dubbed the "gamma boson" referring to the Dirac matrices, $\gamma^\mu$, which are promoted to be the quantum field for gravity at the scale of elementary particles. Instead, the graviton appears merely as a tensor bound state of two gamma bosons in the same spin eigenstate, by referring to the relation $g_{\mu\nu} = \tfrac{1}{2} (\gamma_\mu \gamma_\nu + \gamma_\nu \gamma_\mu)$ and the metric $ds^2 = g_{\mu\nu} dx^\mu dx^\nu = (\gamma_\alpha dx^\alpha)^2$. Consequently, like the electroweak theory and quantum chromodynamics, gravity may be formalized as a Yang-Mills theory. As a consequence, there is no need of the Higgs field or any symmetry breaking mechanism to generate the mass of fundamental particles. We show that one can get rid of the Yukawa couplings in favor of the covariant derivative. Finally, a set of partial differential equations that is equivalent to Einstein equations is established for the gamma boson. Static spherical symmetric external solutions leading to the Schwarzschild metric are found by solving the latter neglecting the mass density of the gravitational field itself. Taking into account the latter involves a departure from the Schwarzschild solution that may be tested by laboratory experiments or astrophysical observations.

## 1 Introduction.

General relativity (GR) through the concept of curved spacetime gives us our best understanding of matter-energy gravity, while quantum field theory (QFT) through the concept of duality field(wave)-particle gives us our best understanding of matter-energy exchange. Both fundamental theories, though very different conceptually from each other, have been tested until now with a great accuracy and impressive precision. Most of the theoretical efforts to reconcile these two fundamental approaches rely on superstring theory [1], loop quantum gravity [2] and to a lesser extend to non-commutative geometry [3]. However, all these attempts cannot yet claim success until now in explaining or predicting any new experimental or observation facts. Even the problem of the quantization of spacetime itself, quantum gravity, is still open. For the meantime, the standard model of particle physics allows to understand very precisely both the electroweak and strong interaction with the same gauge field formalism, namely the Yang-Mills theory based on the compact Lie groups U(1)XSU(2)XSU(3) and vector gauge fields (the photon, the Z and W-bosons, and the gluons). Nevertheless, by requirering the unification of all fundamental interactions within the same mathematical framework, gravity needs to be included. Until now, this has been a difficult task, since gravity is well understood within the framework of general relativity (GR). Now, though GR is proved to be a gauge theory too, it is rather a classical field theory based on the non compact Lie group of diffeomorphisms GL(4, R) and a tensor gauge field, the graviton. Moreover, the Yang-Mills gauge theories are consistent only with zero mass vector bosons. Therefore, unless the existence of the Higgs field is confirmed, we do not understand the origin of either the $Z^0$ and W-bosons masses or the fermions masses. Besides, the standard model meets the problem of the huge vacuum energy which seems in conflict with the observed tiny cosmological constant, in consistency with the accelerating universe [4], derived from the concordance model of cosmology [5]. Clearly, things would be better for fundamental physics, if gravity itself could be described at the level of particle physics as a Yang-Mills theory and if at the same time one could get rid of the Higgs boson and the Higgs mechanism of symmetry breaking. Finally, the resemblances quoted in the litterature between gravity and electromagnetism should be reconsidered in this respect. Indeed, since the failure of the search of a coherent relativistic scalar gravitational theory and the advent of GR, the too numerous differences in mathematical structure between GR (non abelian and non compact gauge theory) and Maxwell theory (abelian and compact gauge theory) are more often emphasized. Let us point out that these differences are accentuated at the macroscopic scale since the "gravitational charge", namely the mass, may get huge as compared to the electric charge which may cancel out, as it is the case for heavenly bodies. In other words, it is still possible that at the level of fundamental particles, the resemblance between gravity and electromagnetism could be much more justified. It is the aim of this paper to explore such a possibility, recalling that gravity is not well tested at distances below 100 μm [6]. More precisely, gravity is reconsidered at the level of elementary particles as an abelian Yang-Mills theory whose tensorial feature manifests itself at the macroscopic scale as a consequence of curved spacetime. As a matter of fact, spacetime is almost flat at the very small scale and it gets curved in the presence of a massive body at the macroscopic scale. In other words, spacetime curvature may be looked at as a macroscopic phenomenon which is quite negligible, up to quantum corrections involving the Planck constant, h, when the source of the gravitational field reduces to an elementary particle. Earlier attempts [7], [8] to treat gravity within the framework of the Yang-Mills theory based respectively on a SU(2)xU(1) gauge group and SO(3, 2) gauge group could recover the Einstein-Hilbert Lagrangian density only in the weak field approximation. Also, these theories still need the Higgs mechanism for the elementary particles acquire their masses.

.
## 2 GR as a U(1) gauge theory.

As one knows, the Dirac equation looks like an eigenvalue equation where the rest energy, $mc^2$, is the eigenvalue of the hermitian operator H = iℏc $\gamma^\mu \partial_\mu$ [9, 10]. Indeed, the Dirac equation of a free fermion reads

iℏ $\gamma^\mu \partial_\mu \psi$ - mc $\psi$ = 0, (1)

and this follows from the Lagrangian density, $L_\psi$, of the fermion described by the Dirac spinor $\psi$,

$$L_\psi = \tfrac{1}{2} i\hbar c\, \underline{\psi}\, \gamma^\mu \partial_\mu \psi - mc^2\, \underline{\psi}\, \psi + \text{h.c.}, \quad (2)$$

where we have set $\underline{\psi} = \psi^+ \gamma^0$. Since $\gamma^\mu \gamma_\mu = 4$, equation (1) and relation (2) can be recasted respectively into the following forms

$$i\hbar\, \gamma^\mu D_\mu \psi = 0 \quad (3)$$

and

$$L_\psi = \tfrac{1}{2} i\hbar c\, \underline{\psi}\, \gamma^\mu D_\mu \psi + \text{h.c.}.\quad (4)$$

By defining the covariant derivative

$$D_\mu = \partial_\mu + i\, q_\psi\, \gamma_\mu, \quad (5)$$

where the charge $q_\psi$ of the fermion for the interaction with the gamma-field is related to its mass m through the relation $q_\psi = mc/4\hbar$. Relation (5) suggests that the Dirac matrices $\gamma^{\mu's}$, as the components of a 4-vector, can be viewed as a U(1) gauge field and that the mass, m, may play the role of the charge. Thus, the mass term in the Lagrangian density (2) can be interpreted as an interaction term. Moreover, the Lagrangian density (2) too implies the conservation of the mass, m, of a free fermion, since it can be associated to the conserved current $J_\mu = i\,(mc^2/4)\, \underline{\psi}\, \gamma_\mu\, \psi$. Clearly such an approach is definitely an alternative to the Higgs field and the Yukawa coupling. As we show further, the outstanding question of the quantization of curved spacetime raised in quantum gravity may find thereby a solution. Now, let us consider the motion of the fermion as a test particle in the curved spacetime of a massive body. Then, the $\gamma^{\mu's}$ become spacetime dependent. As such, according to relation (4), the $\gamma^{\mu's}$ appear as auxiliary fields. Adding a kinetic term for the $\gamma^{\mu's}$ to the Lagrangian density (4) makes them components of a true dynamical vector field.

The form of the latter kinetic term is choosen in analogy to the Lagrangian density of the Maxwell field quantization in the so-called Feynman gauge [11, 12][1,2], this reads

$$L_\gamma = (c^4/16\pi G)\,[\,-\nabla_\mu \gamma^\nu \nabla_\nu \gamma^\mu + (\nabla_\alpha \gamma^\alpha)^2\,]. \quad (6)$$

In addition, let us notice that the above kinetic term is analogous to that of the Horava-Lifshitz gravity theory in 3 + 1 dimensions (up to the Cotton tensor contribution) that satisfies full spacetime diffeomophisms invariance [13], by replacing the extrinsic curvature tensor of the preferred time foliation of spacetime $K_{ij}$ by the covariant derivative $c^3\, \nabla_\mu \gamma^\nu$.

---

[1] Following the usual procedure one would start with the Lagrangian density $L = -\tfrac{1}{4} F_{\mu\nu} F^{\mu\nu}$ where the antisymmetric field strength $F_{\mu\nu} = \nabla_\mu A_\nu - \nabla_\nu A_\mu$ and the 4-vector field $A_\mu$ is subject to the Lorentz gauge $\nabla_\alpha A^\alpha = 0$. However, as one knows, the Lorentz gauge is inconsistent with the canonical commutation relations between the spin one field $A^\mu$ and its momenta $\nabla_\nu A_\mu$. So, one is led to use instead the Feynman gauge by redefining the Lagrangian density $L = -\tfrac{1}{4} F_{\mu\nu} F^{\mu\nu} - \tfrac{1}{2}\xi\,(\nabla_\alpha A^\alpha)^2$ which by choosing $\xi = 1$ (Feynman gauge) defines the Fermi Lagrangian $L = -\nabla_\mu A^\nu \nabla^\mu A_\nu$.

[2] Note that the torsion-free condition $\nabla^\nu \gamma^\mu = \nabla^\mu \gamma^\nu$ (see ref. [14], relation (5.7)) yields the identity $\nabla_\mu \gamma^\nu \nabla_\nu \gamma^\mu = \nabla^\nu \gamma_\mu \nabla_\nu \gamma^\mu$. Besides, $-\nabla_\mu \gamma^\nu \nabla_\nu \gamma^\mu + (\nabla_\alpha \gamma^\alpha)^2 = -\nabla_\mu \gamma^\nu \nabla_\nu \gamma^\mu$ with $\mu \neq \nu$.

Applying the minimal coupling prescription, the covariant derivative (5) rewrites in curved spacetime

$$D_\mu = \nabla_\mu + i\, q_\psi\, \gamma_\mu. \quad (7)$$

So that the total Lagrangian density reads

$$L = L_\gamma + L_\psi = (c^4/16\pi G)\left[-(\nabla_\mu \gamma^\nu)(\nabla_\nu \gamma^\mu) + (\nabla_\alpha \gamma^\alpha)^2\right] + \tfrac{1}{2} i\hbar c\, \overline{\psi}\, \gamma^\mu\, D_\mu \psi. \quad (8)$$

Now, 
$$\begin{aligned}
L_\gamma &= (c^4/32\pi G)\{-[(\nabla_\mu \gamma^\nu)(\nabla_\nu \gamma^\mu) + (\nabla_\nu \gamma^\mu)(\nabla_\mu \gamma^\nu)] + 2(\nabla_\alpha \gamma^\alpha)^2\} \\
&= (c^4/32\pi G)\{[\gamma^\nu \nabla_\mu \nabla_\nu \gamma^\mu - \nabla_\mu(\gamma^\nu \nabla_\nu \gamma^\mu) + (\nabla_\mu \nabla_\nu \gamma^\mu)\gamma^\nu - \nabla_\mu((\nabla_\nu \gamma^\mu)\gamma^\nu)] \\
&\quad - \gamma^\nu \nabla_\nu \nabla_\mu \gamma^\mu + \nabla_\nu(\gamma^\nu \nabla_\mu \gamma^\mu) - (\nabla_\nu \nabla_\mu \gamma^\mu)\gamma^\nu + \nabla_\nu((\nabla_\mu \gamma^\mu)\gamma^\nu). \quad (9)
\end{aligned}$$

Dropping the divergence terms $\nabla_\mu(\gamma^\nu \nabla_\nu \gamma^\mu)$, $\nabla_\mu((\nabla_\nu \gamma^\mu)\gamma^\nu)$, $\nabla_\nu(\gamma^\nu \nabla_\mu \gamma^\mu)$ and $\nabla_\nu((\nabla_\mu \gamma^\mu)\gamma^\nu)$, the Lagrangian density (9) above reduces to

$$\begin{aligned}
L_\gamma &= (c^4/32\pi G)\left[\gamma^\nu(\nabla_\mu \nabla_\nu \gamma^\mu - \nabla_\nu \nabla_\mu \gamma^\mu) + (\nabla_\mu \nabla_\nu \gamma^\mu - \nabla_\nu \nabla_\mu \gamma^\mu)\gamma^\nu\right] \\
&= -(c^4/32\pi G)(\gamma^\nu R_{\mu\nu}{}^{\mu\lambda}\gamma_\lambda + R_{\mu\nu}{}^{\mu\lambda}\gamma_\lambda \gamma^\nu) \\
&= -(c^4/32\pi G)(\gamma^\nu \gamma_\lambda + \gamma_\lambda \gamma^\nu) R_\nu{}^\lambda \\
&= -(c^4/16\pi G)\, \delta^\nu{}_\lambda R_\nu{}^\lambda \\
&= -(c^4/16\pi G)\, R. \quad (10)
\end{aligned}$$

By taking into account that $(\nabla_\alpha \nabla_\beta - \nabla_\beta \nabla_\alpha)\gamma_\mu = -R_{\alpha\beta\mu}{}^\nu \gamma_\nu$ for a torsion-free connection [14] and $g_{\mu\nu} = (\gamma_\mu \gamma_\nu + \gamma_\nu \gamma_\mu)/2$ satisfies the metricity condition $\nabla_\lambda g^{\mu\nu} = 0$.

Therefore, the Lagrangian density (8) becomes

$$L = -(c^4/16\pi G)\, R + \tfrac{1}{2} i\hbar c\, \overline{\psi}\, \gamma^\mu \nabla_\mu \psi - mc^2\, \overline{\psi}\,\psi. \quad (11)$$

where $L_{EH} = -(c^4/16\pi G)\, R$ is the Einstein-Hilbert Lagrangian density.

## 3 Self-interacting gamma-field and the residual cosmological constant.

Since the Euler-Lagrange equations derived from the least action principle applied to the Lagrangian-density (8) are non linear, the gamma-field may be viewed as a self-interacting vector field. So, let us apply the minimal coupling prescription to the Lagrangian density of the gamma-field itself. Then, it follows from relation (6),

$$L_\gamma = (c^4/16\pi G)\left[-(D_\mu \gamma^\nu)^+ D_\nu \gamma^\mu + |D_\alpha \gamma^\alpha|^2\right], \quad (12)$$

where we have made the substitution

$$\nabla_\mu \gamma_\nu \rightarrow D_\mu \gamma_\nu = [\nabla_\mu + i\, q_\gamma\, \gamma_\mu]\, \gamma_\nu, \quad (13)$$

according to the covariant derivative (7) and set $|D_\alpha \gamma^\alpha|^2 = (D_\alpha \gamma^\alpha)^+ (D_\alpha \gamma^\alpha)$. Relation (12) together with relation (13) yield

$$L_\gamma = (c^4/16\pi G) [ - \nabla_\mu \gamma^\nu \nabla_\nu \gamma^\mu + (\nabla_\alpha \gamma^\alpha)^2 + 8 q_\gamma^2 ]$$

$$= - (c^4/16\pi G)(R - 2\Lambda), \quad (14)$$

where $\Lambda = 4 q_\gamma^2$ defines the residual cosmological constant[3]. Let us notice that the self-interacting gamma-field involves a positive cosmological constant but it remains massless.

## 4 Vectorial boson mass acquisition from the interaction with the gamma boson.

Let $L_B = -\frac{1}{2} (F^{\mu\nu})^+ F_{\mu\nu}$ be the Lagrangian density of a non abelian Yang-Mills boson B, where $F_{\mu\nu} = \partial_\mu B_\nu - \partial_\nu B_\mu + i g_B [B_\mu, B_\nu]$ denotes the field strength in flat spacetime and neglecting the interaction of the B-field with the gamma boson. The interaction with the gamma boson is taken into account by applying the minimal coupling prescription,

$$\partial_\mu B_\nu \rightarrow D_\mu B_\nu = [\nabla_\mu + i q_\gamma \gamma_\mu] B_\nu. \quad (15)$$

Thus, the field strength of the B-field rewrites,

$$\mathbf{F}_{\mu\nu} = D_\mu B_\nu - D_\nu B_\mu + i g_B [B_\mu, B_\nu] = F_{\mu\nu} + i q_\gamma (\gamma_\mu B_\nu - \gamma_\nu B_\mu). \quad (16)$$

Expanding and arranging, it follows the Lagrangian density

$$L_B = -\frac{1}{2} (\mathbf{F}^{\mu\nu})^+ \mathbf{F}_{\mu\nu} = -\frac{1}{2} (F^{\mu\nu})^+ F_{\mu\nu} + (m_B c/\hbar)^2 B_\mu^+ B^\mu + L_{int}, \quad (17)$$

where $m_B = \hbar |q_\gamma| \sqrt{2}/c$ is the mass acquired by the B-field by intedracting with the gamma boson and $L_{int} = L_{int}(B^\mu, B^{\nu+}, \nabla_\mu B^\nu, \nabla_\mu B^{\nu+}, \gamma^\alpha, \gamma^{\beta+})$ denotes the corresponding interaction term.

## 5 Equations of the gamma-field and static spherical symmetric solution.

### 5.1 Equations of the gamma-field.

The least action principle $\delta \iiiint (-g)^{1/2} L \, d^3\mathbf{x} \, dt = 0$ applied to the total Lagrangian density (8) yields the Euler-Lagrange equations $\nabla_\nu [\partial L/\partial(\nabla_\mu \gamma_\nu)] = \partial L/\partial \gamma_\mu$, where g is the determinant of the metric tensor. Hence,

---

[3] It is likely that either a symmetry could cancel out all the huge vacuum energy terms derived from QFT or their cancellation could occur dynamically [15]. If so, the self-interacting gamma-field could generate the observed tiny cosmological constant.

$$(c^4/4\pi G) \left[ -\nabla_\nu \nabla^\nu \gamma^\mu + \tfrac{1}{2} (\nabla_\alpha \gamma^\beta \nabla_\beta \gamma^\alpha)_{\beta \neq \alpha} \gamma^\mu \right] = i\hbar c\, \partial (\overline{\psi} \gamma^\alpha D_\alpha \psi)/\partial \gamma_\mu \quad (18)$$

and, on account of equation (1), one gets the following system of partial differential equations

$$-\nabla_\nu \nabla^\nu \gamma^\mu + \tfrac{1}{2} (\nabla_\alpha \gamma^\beta \nabla_\beta \gamma^\alpha)_{\beta \neq \alpha} \gamma^\mu = (4\pi G/c^3)\, i\hbar\, \psi^+ \gamma^0 \nabla^\mu \psi. \quad (19)$$

## 5.2 Static spherical symmetric external solution in the classical limit.

Let us consider a free fermion of mass m as the source of the gamma-field. In the classical limit, that it is for $\hbar \to 0$, an elementary particle endows no curvature to spacetime. Thus, we may solve equations (19) together with the Dirac equation in the flat space approximation. The plane wave solution of the Dirac equation (1) reads

$$\psi = \psi(\mathbf{0}, 0)\, \exp(-i\, P_\mu x^\mu/\hbar). \quad (20)$$

In the rest frame of the particle source which is a fermion, one gets the momentum $P_k = 0$ and $P_0 = mc$, the wave function $\psi = \psi(\mathbf{0}, 0)\, \exp(-i\, P_0 x^0/\hbar)$ and then $\partial_0 \psi = -i\,(mc/\hbar)\, \psi$. Thus equation (19) becomes

$$-\partial_\nu \partial^\nu \gamma_0 + \tfrac{1}{2} (\partial_\alpha \gamma^\beta \partial_\beta \gamma^\alpha)_{\beta \neq \alpha} \gamma_0 = (4\pi Gm/c^2)\, \psi^+ \gamma^0 \psi, \quad (21)$$

$$-\partial_\nu \partial^\nu \gamma^1 + \tfrac{1}{2} (\partial_\alpha \gamma^\beta \partial_\beta \gamma^\alpha)_{\beta \neq \alpha} \gamma^1 = 0. \quad (22)$$

As usual in spherical coordinates $x^0 = ct$, $x^1 = r$, $x^2 = \theta$ and $x^3 = \phi$. Furthermore, assuming static spherical symmetry implies $\partial_\alpha \gamma^\beta = 0$ for $\alpha \neq 1$ and hence $(\partial_\alpha \gamma^\beta \partial_\beta \gamma^\alpha)_{\beta \neq \alpha} = 0$ for $\beta \neq \alpha$. Thus, equations (21) and (22) reduce respectively to

$$\Delta \gamma_0 = (4\pi Gm/c^2)\, \psi^+ \gamma^0 \psi, \quad (23)$$

$$\Delta \gamma^1 = 0. \quad (24)$$

Solving equation (24) on account of the boundary condition $g_{\mu\nu} \to \eta_{\mu\nu}$ for $r \to \infty$, one gets

$$\gamma^1 = [1 - (R_S/r)]\, \gamma^1_\infty, \quad (25)$$

where we have set $\gamma^1(r \to \infty) = \gamma^1_\infty$. Since $g^{11} = \gamma^1 \gamma^1$, it follows $g^{11} = -1 + (R_S/r)$ and hence $g_{11} = (g^{11})^{-1} = -[1 - (R_S/r)]^{-1}$, where $R_S = 2Gm/c^2$. Furthermore, $g_{00} = \gamma_0 \gamma_0$ and $(\gamma^0)^{-1} = \gamma_0$ implies

$$(\Delta \gamma_0)\, \gamma_0 + \gamma_0 \Delta \gamma_0 = \Delta g_{00} + \tfrac{1}{2}\, g^{00}\, (\partial_1 g_{00})\, (\partial^1 g_{00}). \quad (26)$$

Hence, equation (23) becomes

$$\Delta g_{00} + \tfrac{1}{2}\, g^{00}\, (\partial_1 g_{00})\, (\partial^1 g_{00}) = (8\pi G/c^2)\, m\, \psi^+ \psi. \quad (27)$$

By setting $g_{00} = 1 + 2\,(V_N/c^2)$ and $\rho = m\, \psi^+ \psi - (c^2/16\pi G)\, g^{00}\, (\partial_1 g_{00})\, (\partial^1 g_{00})$, equation (27) takes the form of the Poisson's equation

$$\Delta V_N = 4\pi G\, \rho, \quad (28)$$

where $V_N$ denotes the Newtonian potential, $\rho_{mat} = m\psi^+\psi$ defines the mass density of the femion and

$$\rho_G = -(c^2/16\pi G) g^{00} (\partial_1 g_{00})(\partial^1 g_{00}) = g^{00} (\nabla V_N)^2/4\pi Gc^2, \quad (29)$$

is the mass density of the gravitational field itself. Thus, equation (28) includes naturally the mass density, $\rho_G$, of the gravitational field as expected from the equivalence between mass and energy. Equation (28) together with relation (29) are put forward by Rémi Hakim in his text book [16] justifying in this way the need for nonlinear equations for the gravitational field. Since the author deals with a scalar potential, he expresses the mass density of the gravitational field as half of relation (29). Clearly, relation (29) takes into account the two polarization states of a massless vector field, as expected. Besides, the mass density of the gravitational field as expressed in relation (29) is positive[4] not negative as assumed in ref. [16]. Since $\iiint \psi^+\psi \, d^3\mathbf{x} = 1$, neglecting the mass density of the gravitational field, $\rho_G$, the solutions of equations (27) and (28) read respectively $g_{00} = 1 - (R_S/r)$ and $V_N = -Gm/r$.

# 6 Conclusion

Starting from the fact that the Dirac matrices become spacetime dependent in curved spacetime, we have explore the possibility that these be considered as the components of a quantum vectorial field in curved spacetime. Pushing further the analysis in this way, we find that the Einstein-Hilbert action may be derived exactly from the action of this quantum vectorial field, dubbed the "gamma-field". As a consequence, gravity in its turn may be understood within the framework of the Yang-Mills theory based on a compact Lie group as it is the case for the electroweak and strong interactions. The compact Lie group of gravity at the elementary particle level turns out to be the U(1) gauge group. As a byproduct, an alternative to the Higgs mechanism and the Yukawa couplings follows from the covariant derivative defined by the gamma-field. Gravity at the mascroscopic scale, say beyond a nanometer, emerge from gravity at the microscopic scale. Furthermore, the graviton may be viewed as a bound state of two gamma bosons in the same spin eigenstate. Besides, it is shown that the solutions of the gravitational field may be found from the equations of the gamma-field as well as from Einstein equations in the classical limit. The quantization of the gamma-field should follow the canonical formalism of QFT. In addition, the field under consideration being a massless U(1) gauge field, the quantized theory should be renormalizable. The phenomenology of the gamma-field, especially in the vicinity of strong gravitational sources like supermassive black holes or compact stellar objects, will be adressed elsewhere.

---

[4] One finds $m_G = \int \rho_G \, 4\pi r^2 \, dr = 6Gm^2/5Rc^2 = 2U/c^2$ with $V_N = Gm(r^2 - 3R^2)/2R^3$ for $0 \leq r \leq R$ (interior solution) and $V_N = -Gm/r$ for $r > R$ (external solution), and U denotes the gravitational binding energy for a spherical mass m of uniform density ; hence $m_G \ll m$ for $r \geq R \gg R_S$.